\documentclass[a4paper,12pt]{article}

\usepackage[utf8]{inputenc}
\usepackage[T1]{fontenc}
\usepackage{lmodern}

\usepackage[margin=2.5cm]{geometry} 

\usepackage{amsmath,amssymb,amsthm}
\usepackage{graphicx}
\usepackage{microtype}
\usepackage[square,comma,numbers,sort&compress]{natbib}
\usepackage[colorlinks=true, linkcolor=blue, citecolor=blue, urlcolor=blue]{hyperref}

\title{\textbf{Holographic connection of $f(G)$ gravity through Barrow and a generalized version of holographic dark fluid}}
\author{Surajit Chattopadhyay\\Department of Mathematics, Amity University, Kolkata, India \\ \texttt{schattopadhyay1@kol.amity.edu}}
\date{\today}

\begin{document}
\maketitle

\begin{abstract}
In the context of $f (G)$ modified gravity, we address the cosmic application of the most generalized form of holographic dark energy (\textit{The European Physical Journal C} \textbf{77} (2017): 1-8) in this study, as well as a specific instance of it in the form of Barrow holographic dark energy (\textit{Physical Review D}, \textbf{102}(12), p.123525). Holographic dark energy and a well-known power law form of the scale factor $a(t)$ are added to the $f (G)$ model in order to achieve this. It is observed that a sufficient criterion for a realistic modified gravity model is satisfied by the reconstructed $f (G)$. The reconstruction models are also tested under the four energy situations. \\\\

\textbf{keywords}:$f (G)$ gravity, Barrow holographic dark energy, Generalized holographic dark energy, holographic reconstruction, energy conditions.
\end{abstract}

\section{Introduction}
Due to numerous discoveries, the universe's accelerated expansion is now widely accepted \cite{astier2012expansion, perlmutter1999supernovae, schmidt2012nobel, riess1998observational, riess2020expansion}. The "standard model of cosmology" that scientists currently use is called $\Lambda$ CDM, and it was economically constructed using six free parameters and several tried-and-true ansatzes \cite{riess2020expansion}. Accelerated expansion, structure development, primordial nucleosynthesis, flat spacetime geometry, Big Bang afterglow fluctuations, and the initial combining of baryons into atoms are only a few of the phenomena that the model describes \cite{ riess1998observational, riess2020expansion}. The most accurate estimate of two modern quantities is provided by the ESA Planck satellite's CMB observations, which offer the best current calibration of the parameters in $\Lambda$ CDM. These parameters are used to determine the current Hubble constant and the current age of the Universe \cite{riess1998observational}. According to recent astronomical data, which includes high redshift surveys of supernovae and measurements from the Wilkinson Microwave Anisotropy Probe, a peculiar cosmic fluid with negative pressure, known as "dark energy" (DE), is responsible for almost $70\%$ of the universe's total energy. On the other hand, the universe is currently expanding faster.  Phantom DE is the term used to describe the situation where EoS, denoted as $w$ is smaller than $-1$. Phantom dark energy currently lacks a good theoretical explanation \cite{sahni20045,sahni2003theoretical,odintsov2019unification}. Some significant reviews on DE are available in the works including \cite{copeland2006dynamics, li2013dark, frieman2008dark} and \cite{bamba2012dark}.

The holographic principle \cite{bousso2000holographic, bousso2002holographic, hooft2001holographic}, which has its roots in string theory and black hole thermodynamics, connects the maximum distance of a quantum field theory to its infrared cutoff, which is correlated with the vacuum energy \cite{nojiri2017covariant, nojiri2020unifying, nojiri2021different, nojiri2022barrow}. In cosmology, such holographic consideration is widely used, particularly for depicting the late time dark energy epoch, also referred to as holographic dark energy (HDE) \cite{nojiri2022barrow}. The holographic concept was also very successful in the early universe for both the inflation and bounce scenarios, in addition to describing the dark energy epoch. In conjunction with the dimensional analysis, the holographic principle gives rise to the holographic energy density in formulating holographic cosmology \cite{nojiri2022barrow}. Because of this, holographic cosmology differs greatly from the typical inflation or dark energy models, in which the appropriate agent is represented by one or more suitable scalar fields or higher curvature terms in the Lagrangian \cite{nojiri2022barrow, elizalde2005dark}. The fact that the entropy of the entire universe, viewed as a system with radius, the previously indicated maximum distance, is proportionate to its area, much like a black hole, is a key component of the cosmic application of holography \cite{saridakis2018holographic}. However, Gibbs had already noted in 1902 that the Boltzmann-Gibbs theory could not be applied to systems where the partition function diverges, and we now know that gravitational systems fall into this category \cite{saridakis2018holographic, nojiri2022nonextensive}. According to the gravity-thermodynamic postulate, HDE depicts our universe as a hologram, with Bekenstein-Hawking entropy encoding its degrees of freedom. However, it does not accurately chart the evolution of the universe, which is why appropriate changes should be made \cite{luciano2023saez,luciano2024kaniadakis}. According to the gravity-thermodynamic postulate, HDE depicts our universe as a hologram, with Bekenstein-Hawking entropy encoding its degrees of freedom \cite{luciano2023saez}. However, it does not accurately describe the evolution of the universe, which is why appropriate changes should be made \cite{luciano2023saez}. Accordingly, HDE with deformed horizon entropies—such as Tsallis \cite{tavayef2018tsallis, pandey2022new, chokyi2024cosmology, dheepika2022tsallis}, Kaniadakis \cite{drepanou2022kaniadakis, rani2022cosmographic, odintsov2023holographic}, and Barrow entropies \cite{odintsov2023holographic}—offers a plausible paradigm. These entropies result from attempts to add relativistic, quantum gravity, and non-extensive corrections to the classical Boltzmann-Gibbs statistics, respectively \cite{luciano2023saez}. Having obtained the generalized entropies, the immediate next task is to check their viability, i.e whether the entropic cosmology corresponding to such generalized entropies lead to the correct evolution of universe consistent with the observational data \cite{odintsov2023holographic}. Assuming ordinary matter scales canonically, \cite{guberina2005generalized} demonstrated that the continuity equation unambiguously determines the IR cutoff when a law of variation for the energy density or Newton constant is known. In the generalized HDE formalism proposed by \cite{nojiri2021different}, the holographic cut-off can be expressed as $L_{IR}=L_{IR}(L_p,\dot{L}_P, \ddot{L}_P,...,L_f,\dot{L}_f, \ddot{L}_f,...,a) $, where $L_p$ and $L_f$ represent the particle and future horizons, respectively, where $a$ represents the universe's scale factor.

As one major aspect of the current work is $f(G)$ gravity, which is a candidate of modified gravity, let us have a discussion on some relevant literature. Back in 2004, \cite{nojiri2004modified, nojiri2007modified, nojiri2007modified1} proposed to take into consideration the $ln R$ terms in modified gravity caused by quantum effects in curved spacetime in order to look for a realistic modified gravity that would offer a gravitational substitute for dark energy. With particular reference to modified gravity theories such as $F(R)$, $F(G)$, and string-inspired gravities, including their FRW equations and fluid or scalar-tensor description, \cite{nojiri2011unified, nojiri2014accelerating} showed how modified gravity could be used to describe the early-time inflation and late-time cosmic acceleration of the universe.  \cite{nojiri2007modified} developed a reconstruction scheme for modified $f(R)$ gravity with dark matter, baryons, and radiation by taking into account two versions of such theory. The first one is reconstructed from exact $\Lambda$CDM cosmology, and the second one describes the acceleration era, deceleration–acceleration transition, and sequence of radiation and matter domination. Another study \cite{nojiri2007modified} showed how to explicitly and successfully rebuild some versions from all of those modified gravities using the classical universe expansion history. These modified gravities include $f(R)$, $F(G)$, and string-inspired, scalar-Gauss-Bonnet gravity. They \cite{nojiri2007modified} showed that the acceleration period, deceleration-acceleration transition, and cosmological sequence of matter domination may all appear as cosmological solutions of such a theory. With particular attention to $F(R)$ gravity, \cite{odintsov2018modified} reviewed modified gravity models that provide the unification of inflation with the dark energy era. He also indicated that the gravitational alternative for accelerated expansion is universal, as the corresponding unification may also be achieved in modified Gauss-Bonnet gravity, non-local gravity, or modified gravity with extra scalars. Gauss Bonnet (GB) gravity, also known as $f(G)$ gravity, is one of the many modified gravity candidates that have been proposed in the literature. It is a topological invariant in four dimensions of spacetime, with $G = R- 4R_{\mu \nu} R^{\mu \nu} + R_{\mu \nu \alpha \beta}R^{\mu \nu \alpha \beta}$ \cite{bamba2012dark}. Either $f(G)$ must be an arbitrary function of $G$, or the equation of motion for this gravity must be connected with a scalar field. This modified gravity theory could benefit from the study of the inflationary era, the change from deceleration to acceleration regimes, passing solar system experiment-induced tests, and crossing the phantom divide for various feasible $f(G)$ models \cite{nojiri2002friedmann}. In terms of e-folding, or redshift, \cite{nojiri2009cosmological} stated that a cosmological reconstruction scheme for changed gravity has been created. It is shown how any particular theory can lead to any FRW cosmology. \cite{nojiri2009cosmological} described specific instances of well-known cosmological development that have been recreated, such as $\Lambda$CDM cosmology and deceleration with transition to the phantom super acceleration era, which may or may not generate singularity.

Given the literature surveyed above, we intend to report a reconstruction scheme for $f(G)$ gravity through two different versions of holographic dark energy and a power law form of the scale factor. The remaining part of the paper is organized as follows: Section 2 elaborates on the reconstruction cosmology of $f(G)$ gravity by considering a holographic connection through Barrow holographic dark energy. Section 3 demonstrates connectivity between $f(G)$ gravity and the Nojiri-Odintsov holographic cutoff. In Section 4, we have discussed the relation of the model to observational bounds. We have concluded in Section 5.

\section{Barrow holographic connection with \texorpdfstring{$f(G)$}{f(G)} gravity}
Modified gravity offers a natural gravitational alternative for dark energy, as \cite{nojiri2007introduction} showed. Some sub-dominant factors (such as $1/R$) may become essential with tiny curvature since they detail in detail how the expansion of the cosmos explains the cosmic speed-up. Additionally, they \cite{nojiri2007introduction} discussed how the distinct roles of gravitational components applicable at small and large curvature offer a very natural unification of early-time inflation and late-time acceleration under modified gravity. In the late universe, \cite{nojiri2007introduction} demonstrated how various forms of $f(R)$, $f(G)$, or $f(R, G)$ gravity result in cosmic speed-up by examining models with non-linear gravitational coupling or string-inspired models with Gauss-Bonnet-dilaton coupling. Due to gravitational terms that increase as scalar curvature decreases, some of these theories may naturally describe the effective (cosmological constant, quintessence, or phantom) late-time era with a potential transition from deceleration to acceleration \cite{nojiri2007introduction}. These theories may also pass the Solar System tests and have a rich cosmological structure. In 2005, \cite{nojiri2005modified} proposed a novel type of modified gravity in which the function $G$, which is Gauss-Bonnet (GB) invariant, modifies Einstein's action. Here $G$ is known to be topologically invariant in four dimensions, but in a higher-dimensional brane-world approach, it could result in various intriguing cosmological implications. They \cite{nojiri2005modified} showed that the transition from deceleration to acceleration, crossing the phantom divide, and current acceleration with effective (cosmological constant, quintessence, or phantom) equation of the state of the universe are some of the most intriguing aspects of late-time cosmology that may be described by modified GB gravity. In another notable study, \cite{odintsov2016gauss} used higher-order terms that contained the partial derivative of the Gauss-Bonnet scalar coupled to the baryonic current to discuss some variant forms of gravitational baryogenesis. They \cite{odintsov2016gauss} demonstrated that this scenario extends the well-known theory that uses a similar coupling between the Ricci scalar and the baryonic current.  

The kind of connectivity between modified gravity and HDE is not new in the literature. The literature has previously discussed the connection between HDE and modified gravity. \cite{setare2008holographic} employed the holographic model of dark energy to determine the equation of state for the holographic energy density in a spatially flat universe in order to study the cosmic application of HDE density in the modified gravity framework. In the spatially flat FRW universe, \cite{karami2011reconstructing} reconstructed the various $f(R)$ modified gravity models using the entropy-corrected and ordinary versions of the new agegraphic and holographic dark energy models, which characterize the universe's accelerated expansion. In order to create a modified gravity action that is consistent with the HDE scenario, the work of \cite{chattopadhyay2013reconstruction} used HDE to design a reconstruction strategy for the modified gravity with $f(T)$ action. \cite{devi2024barrow}  determined the Hubble parameter and universe's scale factor, solving the model's field equations and reconstructing the Barrow holographic dark energy within the framework of $f(R,T)$ gravity while taking into account a flat Friedmann-Lemaitre-Robertson-Walker line element. While the HDE model has been covered in the theory, the primary goal of the paper is to discuss the BHDE model within the context of $f(G)$ gravity. Several works have examined the BHDE model in the context of GR, convincingly demonstrating its recent accelerated expansion. However, as an observational examination in the paper by \cite{anagnostopoulos2020observational} shows, issues such as $H_0$ tension still exist with this DE model in GR. 

\subsection{An overview of $f(G)$ gravity}
This section will outline some of the key characteristics of $f(G)$ gravity and show how $f(G)$ and HDE are related. To do this, we look at an action representing a particular kind of the $f(G)$ gravity model \cite{de2009construction, myrzakulov2011lambda, jawad2013reconstruction}, which has an arbitrary Gauss-Bonnet function and an Einstein gravity term with perfect fluid. There is a de Sitter point in the framework of $f(G)$ gravity that can be utilized for cosmic acceleration. The model with inverse powers of linear combinations of quadratic curvature invariants was demonstrated in the work of \cite{de2007unsuccessful}. 
The action $S$ of this $f(G)$ modified
gravity is given by
\begin{equation}\label{1}
S=\int
d^4x\sqrt{-g}\left[\frac{1}{2\kappa^{2}}R+f(G)+\mathcal{L}_{m}\right],
\end{equation}
where
$G=R^{2}-4R_{\mu\nu}R^{\mu\nu}+R_{\mu\nu\lambda\sigma}R^{\mu\nu\lambda\sigma}$
(with $R$ representing the Ricci scalar curvature, $R_{\mu\nu}$ is
the Ricci curvature tensor and $R_{\mu\nu\lambda\sigma}$ denotes the
Riemann curvature tensor), $\kappa^{2}=8\pi G_N$ (with $G_N$ being
the gravitational constant), $g$ represents the determinant of the
metric tensor $g_{\mu \nu}$ and $\mathcal{L}_{m}$ represents the
Lagrangian of the matter present in the universe. The variation of
$S$ with respect to the metric tensor $g_{\mu\nu}$ generates the
field equations. For spatially flat FRW metric, the Ricci scalar
curvature $R$ and the Gauss-Bonnet invariant $G$, take the following
expressions respectively
\begin{equation}\label{2}
R=6(\dot{H}+2H^{2}), \quad G=24H^{2}(\dot{H}+H^{2}),
\end{equation}
where the upper dot represents the derivative with respect to the
cosmic time $t$.

The first FRW equation (with $8 \pi G_N=1)$ takes the form
\begin{equation}
H^{2}=\frac{1}{3}\left[Gf_{G}-f(G)-24\dot{G}H^{3}f_{GG}+\rho_{m}\right]=\frac{1}{3}\left(\rho_{G}+\rho_{m}\right),
\label{4}
\end{equation}
where $f_G$ and $f_{GG}$ represents, respectively, the first and the
second derivative of $f$ with respect to $G$, i.e.
$f_G=\frac{df}{dG}$ and $f_{GG}=\frac{d^2 f}{dG^2}$. \cite{setare2008holographic}
recently reconstructed $f(R)$ modified gravity from HDE with IR
cutoff as the event horizon.  In the FRW universe,
the energy conservation law can be expressed as the standard
continuity equation
\begin{equation}
    \dot{\rho_G}+3H\rho_G(1+w_G)=0
\end{equation}
\begin{equation}
    \dot{\rho_m}+3H \rho_m=0
\end{equation}
where, $w_G$ denotes the EoS parameter due to $f(g)$ gravity. We choose our scale factor $a(t)=a_0 t^\gamma$ leading to $H=\frac{\gamma}{t}$. Hence from Eq. (\ref{2}) we obtain the following form for $G$ in terms of $t$
\begin{equation}\label{G}
G=\frac{24(-1+\gamma)\gamma^3}{t^4}
\end{equation}
and its time derivative comes out to be
\begin{equation}\label{Gdot}
\dot{G}=-\frac{96(-1+\gamma)\gamma^3}{t^5}
\end{equation}
In order to cure finite-time future singularities that arise in the late-time cosmic accelerating epochs, \cite{bamba2012dark} took into consideration specific actual $f (G)$ configurations. Next, they \cite{bamba2012dark} used the most recent estimated numerical values of the deceleration, Hubble, snap, and jerk parameters to develop the viability bounds of these models generated by weak and null energy conditions. In the current study, we are intending to demonstrate a holographic connectivity of $f(G)$ gravity, first through Barrow holographic dark energy (BHDE) and next through a highly generalized holographic dark energy with Nojiri-Odintsov cutoff. In the following subsection, we present an overview of the BHDE.
\subsection{An overview of BHDE}
When applying the holographic principle to a cosmological framework, it is crucial to remember that, like the Bekenstein-Hawking entropy of a black hole, the universe horizon, or maximum distance, entropy is proportional to its area \cite{saridakis2020barrow}. Lately, however, Barrow has demonstrated how complex, fractal characteristics could be introduced into the black-hole structure via quantum gravity phenomena, motivated by the COVID-19 virus drawings \cite{saridakis2020barrow}. Due to this intricate structure, the black-hole entropy expression is distorted and has a finite volume but an infinite (or finite) area \cite{saridakis2020barrow, barrow2020area}:
\begin{equation}\label{barrowentropy}
    S_B=\left(\frac{A}{A_0}\right)^{1+\frac{\Delta}{2}}
\end{equation}
where $A$ is the standard horizon area and $A_0$ the Planck area. Thus, the new exponent $\Delta$ quantifies the quantum-gravitational deformation. In case $\Delta=0$, we obtain the standard Bekenstein-Hawking entropy, and $\Delta=1$ implies the most intricate fractal behaviour \cite{saridakis2020barrow}. Instead of employing the standard Bekenstein-Hawking relation for the horizon entropy, Saridakis \cite{saridakis2020barrow} used the extended Barrow relation to generate the Barrow holographic dark energy (BHDE). In the $\Delta=0$ scenario, Barrow holographic dark energy has the same limit as ordinary holographic dark energy. \cite{saridakis2020barrow} did, however, generally create a new scenario with more complex structure and cosmic behavior. The BHDE proposed by \cite{saridakis2020barrow} would be taken into consideration in our current investigation of the Barrow holographic relationship with $f(G)$ gravity. Under the imposition $S \propto A \propto L^2$ \cite{saridakis2020barrow, wang2017holographic}, the standard holographic dark energy is determined by the condition $\rho_{DE}L^4 \le S$, where $L$ is the horizon length. Using Eq. (\ref{barrowentropy}), the density of BHDE is \cite{saridakis2020barrow}:
\begin{equation}\label{BHDE}
\rho_{\Lambda}=\mathbb{C} L^{\Delta-2}
\end{equation}
where, $\mathbb{C}$ a parameter with dimensions$[L]^{-2-\Delta}$. In case $\Delta=0$, the above expression gives us back the standard holographic dark energy. 

We consider a flat FRW metric
\begin{equation}\label{FRW}
ds^2=-dt^2+a^2(t)\delta_{ij}dx^i dx^j
\end{equation}
In this case, the scale factor is $a(t)$. While there are other options for determining the maximum length $L$ that can be found in the expression of any holographic dark energy, the future event horizon is the one that is most frequently used in the literature \cite{saridakis2020barrow, li2004model} and is given by
\begin{equation}\label{futurevent}
R_h \equiv a\int_t^{\infty}\frac{dt}{a}=a\int_a^{\infty}\frac{da}{Ha^2}
\end{equation}
where, $H=\frac{\dot{a}}{a}$. Replacing $L$ by $R_h$ in Eq.(\ref{BHDE}) we obtain BHDE density as
\begin{equation}\label{BHDERh}
\rho_{\Lambda}=\mathbb{C} R_{h}^{\Delta-2}
\end{equation}
For the choice of the scale factor mentioned in the previous section, we get the form of $R_h$ as
\begin{equation}\label{Rh}
R_h=\frac{t}{-1+\gamma}+C_1 t^\gamma
\end{equation}
and using Eq. (\ref{Rh}) in (\ref{BHDERh}) we obtain
\begin{equation}\label{BHDE}
\rho_{\Lambda}=\mathbb{C} \left(C_1 t^{\gamma }+\frac{t}{-1+\gamma }\right)^{-2+\Delta }    
\end{equation}
Using binomial expansion and some truncation, we get the following form of BHDE density with the power-law scale factor already mentioned: 
\begin{equation}\label{BHDEapprox}
\rho_{\Lambda}=\mathbb{C} \left(\frac{t}{-1+\gamma }\right)^{-2+\Delta } \left(1+\frac{C_1 (-1+\gamma ) (-2+\Delta )}{t}\right)    
\end{equation}
In the following subsection, we would demonstrate the BHDE reconstruction of $f(G)$ gravity. With a similar methodology, this reconstruction scheme, to be elaborated in the subsequent section,  might be comparable to the model investigated by \cite{elizalde2010unifying}.  They \cite{elizalde2010unifying} investigated the modified $f (R)$ Horava-Lifshitz gravity and offered a cohesive explanation of the theory's early-time inflation and late-time acceleration. 
\subsection{The reconstruction scheme}
In this section, we intend to reconstruct $f(G)$ gravity by considering the BHDE connection. For this purpose, we consider $\rho_{\Lambda}=\rho_G$ and hence using (\ref{4}) and (\ref{BHDEapprox}) we get the following differential equation with cosmic time $t$ as the independent variable. 
\begin{equation}\label{ode}
 f(t)+\frac{1}{4} t \frac{df(t)}{dt}-\frac{\gamma ^3 \left(5 \frac{df(t)}{dt}+t \frac{d^2f(t)}{dt^2}\right)}{2 t^3}=-\mathbb{C}\left(\frac{t}{-1+ \gamma }\right)^{-2+\Delta }\left(1+\frac{(-2+\Delta ) C_1 (-1+ \gamma )}{t}\right)  
\end{equation}
In the above equation, we have used Eqs. (\ref{G}) and (\ref{Gdot}) also. The solution would give us reconstructed $f(G)$ as a function of $t$. Considering $\gamma=2$, we get the solution for reconstructed $f(G)$ as 

\begin{equation}\label{freconst}
\begin{aligned}
f(G(t)) &= \frac{1}{(1+\Delta)(2+\Delta)} \, t^{-(6+\Delta)} \Bigg[
\left(t^4\right)^{\frac{2+\Delta}{4}} 
\Big(-4 C_1 t^{1+\Delta} \mathbb{C}(-4+\Delta^2) \\
&\quad + (1+\Delta)\Big(-4 t^{2+\Delta}\mathbb{C} + (2+\Delta)\big(C_2 + 16 e^{t^4/64} C_3 \big)\Big)\Big) \\
&\quad - 2^{\frac{3(1+\Delta)}{2}} C_1 e^{t^4/64} t^{1+\Delta} \left(t^4\right)^{1/4} \mathbb{C}(-4-4\Delta+\Delta^2+\Delta^3) \\
&\quad \times \Gamma\Big[\frac{1+\Delta}{4}, \frac{t^4}{64}\Big] 
- 2^{3+\frac{3\Delta}{2}} e^{t^4/64} t^{2+\Delta} \mathbb{C}(2+3\Delta+\Delta^2) \\
&\quad \times \Gamma\Big[\frac{2+\Delta}{4}, \frac{t^4}{64}\Big]
\Bigg]
\end{aligned}
\end{equation}

that leads to
\begin{equation}\label{fGreconst}
    \begin{array}{c}
f_G(t)=\frac{1}{768 t^2 (-1+\gamma ) \gamma ^3 (1+\Delta ) (2+\Delta )}\left(t\right)^{ -(1+\Delta )} \\
\left(2^{\frac{1}{2}+\frac{3
\Delta }{2}} C_1 e^{\frac{t^4}{64}} t^{3+\Delta } \left(-64+t^4\right) \mathbb{C} \left(-4-4 \Delta +\Delta ^2+\Delta ^3\right) \Gamma\left[\frac{1+\Delta
}{4},\frac{t^4}{64}\right]+4 t \left(2 t^2 \left(t^4\right)^{\Delta /4}\right.\right.\\
\left.\left.\left(4 C_2 \left(2+3 \Delta +\Delta ^2\right)-C_3
e^{\frac{t^4}{64}} \left(-64+t^4\right) \left(2+3 \Delta +\Delta ^2\right)-16 t^{1+\Delta } \mathbb{C} \left(t (1+\Delta )+C_1 \left(-4+\Delta
^2\right)\right)\right)+\right.\right.\\
\left.\left.2^{3 \Delta /2} e^{\frac{t^4}{64}} t^{\Delta } \sqrt{t^4} \left(-64+t^4\right) \mathbb{C} \left(2+3 \Delta +\Delta ^2\right)
\Gamma\left[\frac{2+\Delta }{4},\frac{t^4}{64}\right]\right)\right)
    \end{array}
\end{equation}
and
\begin{equation}\label{fGGreconst}
    \begin{array}{c}
         f_{GG}(t)=\frac{t^{9-\Delta} }{589824 \sqrt{2} (-1+\gamma )^2 \gamma ^6}\times\\
         \left(-2^{3 \Delta /2} C_1 e^{\frac{t^4}{64}} t^{\Delta } \left(t^4\right)^{3/4} \mathbb{C}
(-2+\Delta ) \Gamma\left[\frac{1+\Delta }{4},\frac{t^4}{64}\right]+\right.\\
\left.2 \sqrt{2} \left(2 \left(t^4\right)^{\Delta /4} \left(C_3e^{\frac{t^4}{64}}
t^3+4 t^{\Delta } \mathbb{C} (t+C_1 (-2+\Delta ))\right)-2^{3 \Delta /2} e^{\frac{t^4}{64}} t^{1+\Delta } \sqrt{t^4} \mathbb{C} \Gamma\left[\frac{2+\Delta
}{4},\frac{t^4}{64}\right]\right)\right)
    \end{array}
\end{equation}
As we have reconstructed $f(G)$, $f_G$ and $f_{GG}$, we can have reconstructed $\rho_G$ and considering the conservation equation we have the reconstructed EoS parameter as follows:
\begin{equation}\label{wG}
\begin{aligned}
w_G &= -1 - 
\dfrac{
\begin{aligned}[t]
& t \left(t^4\right)^{1/4} \Bigg[ 
- 2^{3 \Delta/2} C_1 e^{t^4/64} t^{4+\Delta} \mathbb{C} (80 + t^4 -16 \gamma)(-2+\Delta) 
\Gamma\Big[\frac{1+\Delta}{4}, \frac{t^4}{64}\Big] \\
&\quad + 2 \sqrt{2} \left(t^4\right)^{1/4} \Big( 
2 \left(t^4\right)^{\Delta/4} \Big( C_3 e^{t^4/64} t^3 (80+t^4-16\gamma) \\
&\qquad + 4 t^\Delta \mathbb{C} \Big(C_1(-2+\Delta)(t^4 + 16 (2-\gamma + \Delta)) + t(t^4 +16 (3-\gamma+\Delta))\Big) \Big) \\
&\quad - 2^{3\Delta/2} e^{t^4/64} t^{1+\Delta} \sqrt{t^4} \mathbb{C} (80+t^4-16\gamma) 
\Gamma\Big[\frac{2+\Delta}{4}, \frac{t^4}{64}\Big] \Big) 
\Bigg]
\end{aligned}
}{
\begin{aligned}[t]
& 24 \sqrt{2} \gamma \Bigg[
- 2^{1/2 + 3\Delta/2} C_1 e^{t^4/64} t^{1+\Delta} \left(t^4\right)^{1/4} \mathbb{C} (16+t^4-16\gamma)(-2+\Delta)
\Gamma\Big[\frac{1+\Delta}{4}, \frac{t^4}{64}\Big] \\
&\quad + 4 \Big( 2 \left(t^4\right)^{(2+\Delta)/4} \Big( C_3 e^{t^4/64} (16+t^4-16\gamma) + 4 t^{1+\Delta} \mathbb{C} (t + C_1(-2+\Delta)) \Big) \\
&\quad - 2^{3\Delta/2} e^{t^4/64} t^{2+\Delta} \mathbb{C} (16+t^4-16\gamma) 
\Gamma\Big[\frac{2+\Delta}{4}, \frac{t^4}{64}\Big] \Big) 
\Bigg]
\end{aligned}
}
\end{aligned}
\end{equation}

\begin{figure}
    \centering
    \includegraphics[width=0.55\linewidth]{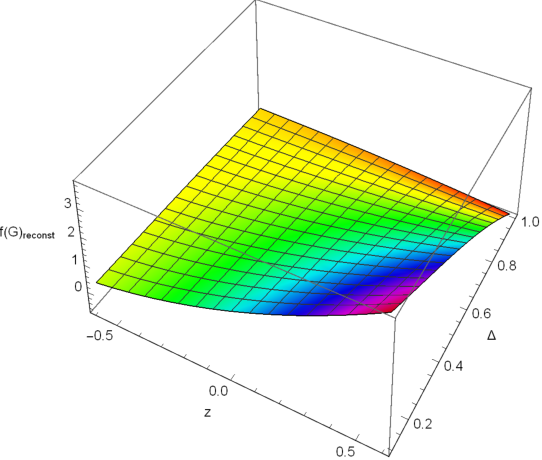}
    \caption{Evolution of Barrow holographically reconstructed $f(G)$ with redshift for a range of values of $\Delta$. The corresponding expression for $f(G)$ is obtained in Eq. (\ref{freconst}). In this plot, we have used $C_1 = 1.5, ~C_2 = 1.3,$ $ ~C_3=0.1,~\gamma=2, ~\mathbb{C}=0.8$.}
    \label{fig1}
\end{figure}
First, let us have a look into Fig. \ref{fig1}, where the Barrow holographically reconstructed $f(G)$ gravity is plotted for a change in $\Delta$. We previously explained that the standard holographic dark energy is established by the condition $\rho_{DE}L^4 \le S$, where $L$ is the horizon length and in the case $\Delta=0$, the BHDE density returns the standard holographic dark energy under the imposition $S \propto A \propto L^2$ \cite{saridakis2020barrow, wang2017holographic}. When charting the reconstructed $f(G)$ in Fig. \ref{fig1}, where the evolutionary pattern of $f(G)$ has been examined against redshift $z$, we have taken this into consideration and limited ourselves to $\Delta \in (0,1)$. It is evident from Fig. 1 that the reconstructed $f(G)$ depends on both $z$ and $\Delta$. First, the reconstructed $f(G)$ exhibits an increasing pattern with $z$. This shows that the reconstructed $f(G)$ is decaying as the universe evolves. Second, the decay is more flattening for $\Delta \approx 1$ and more acute for smaller values of $\Delta$.

\begin{figure}
    \centering
    \includegraphics[width=0.55\linewidth]{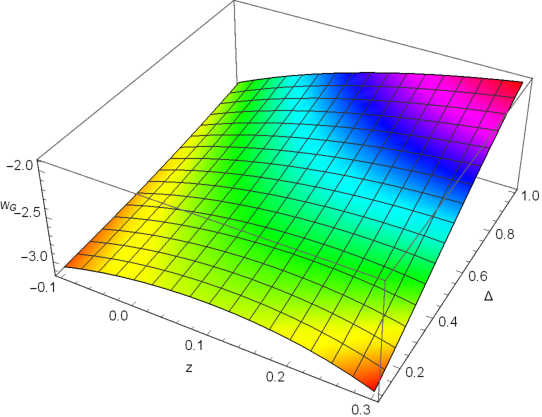}
    \caption{Evolution of reconstructed EoS parameter $w_G$ as obtained in Eq. (\ref{wG}). }
    \label{fig2}
\end{figure}

In Fig. \ref{fig2} we plotted the reconstructed EoS parameter for Barrow holographic $f(G)$ gravity for a range of values of $\Delta$. It is observed that the EoS parameter is staying below $-1$, and hence we can say that it is behaving like a phantom.  Numerous authors with diverse backgrounds have worked on energy conditions (EC). We shall examine the well-known ECs in this work to verify the model's applicability to cosmic acceleration. For the substance of the Universe, various ECs are provided, including the weak energy conditions (WEC), null energy conditions (NEC), dominant energy conditions (DEC), and strong energy conditions (SEC). It is clear from Fig. \ref{fig1} that the reconstructed $f(G)$ depends on both $z$ and $\Delta$ and Fig. \ref{fig2} shows that although the EoS parameter depends upon $z$. It also varies significantly with $\Delta$. The decaying pattern of $w_G$ with the evolution of the universe is more sharp with higher values of $\Delta$ than smaller ones. It is The ECs are the following \cite{koussour2022quintessence}:
\begin{enumerate}
    \item WEC: $\rho>0$
    \item NEC: $p+\rho \ge 0$
    \item DEC: $\rho-p \ge 0$
    \item SEC: $\rho+3p \ge 0$
\end{enumerate}
\begin{figure}[ht] 
  \begin{minipage}[b]{0.55\linewidth} 
    \includegraphics[width=.75\linewidth]{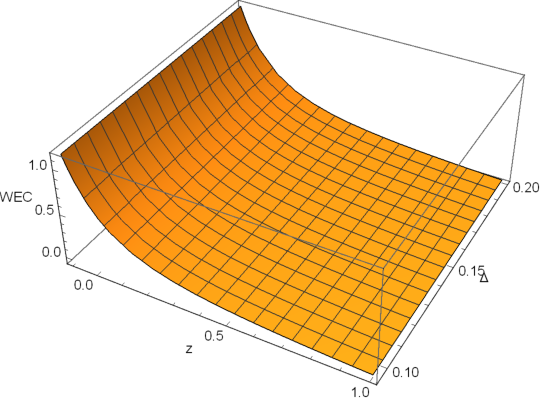} 
    \caption{$\rho_G \ge 0$.}\label{EC}  
  \end{minipage} 
  \begin{minipage}[b]{0.55\linewidth}
    \includegraphics[width=.75\linewidth]{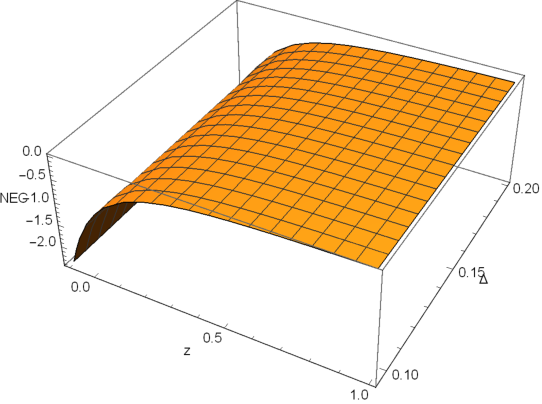} 
    \caption{$\rho_G+p_G \le 0$.}
  \end{minipage} 
  \begin{minipage}[b]{0.55\linewidth}
    \includegraphics[width=.75\linewidth]{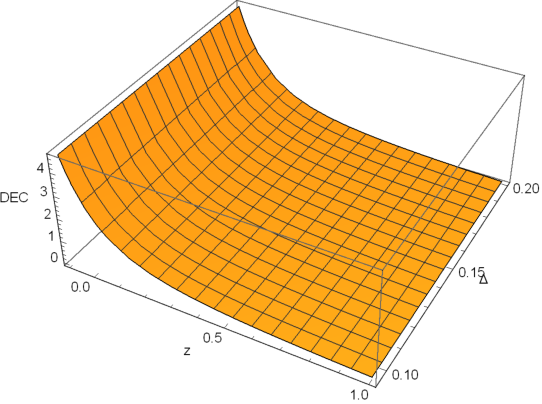} 
    \caption{$\rho_G-p_G \ge 0$}
  \end{minipage}
  \hfill
  \begin{minipage}[b]{0.55\linewidth}
    \includegraphics[width=.75\linewidth]{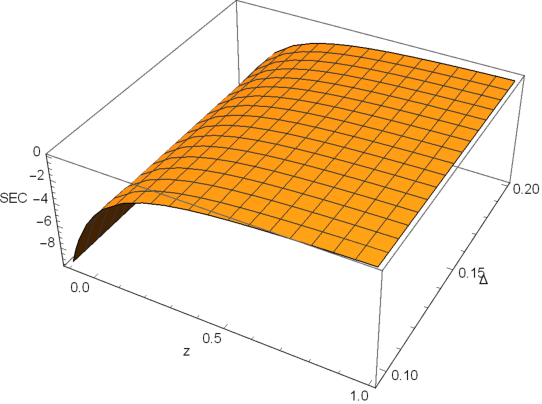} 
    \caption{$\rho_G+3p_G \le 0$}
  \end{minipage} 
\end{figure}
Violation of the NEC represents the depletion of energy density as the Universe expands. In the present case, Figs. 3, 4, 5, and 6 show the ECs of the present model. Fig. 4 shows the violation of NEC. This means $\rho_G+p_G<0$ implying $w_G<-1$. That indicates the possibility of phantom behavior of the EoS parameter. Fig. 6 shows the violation of SEC, i.e. $\rho_G+3 p_G <0$ that implies $w_G<-1/3$. This is consistent with the accelerated expansion of the universe. Fig. 3 shows the validity of WEC, and Fig. 5 shows that DEC is also satisfied.   
\section{Holographic connectivity of $f(G)$ gravity through Nojiri-Odintsov cutoff}
Reference \cite{odintsov2019unification} has demonstrated that the coincidence problem can be overcome in generalized holographic dark energy. In \cite{odintsov2019unification,odintsov2023holographic}, it was postulated that the holographic cut-off (LIR) is influenced by particle and future event horizons, including their derivatives \cite{nojiri2022barrow} 
\begin{equation}\label{23}
    L_{IR} = L_{IR}\left(L_p, \dot{L}_p, \ddot{L}_p,··· , L_f, \dot{L}_f,··· ,a\right)
\end{equation}
Here, $L_p$ and $L_f$ are the particle horizon and future
horizon, respectively, and $a$ is the scale factor of the universe. Using this formalism, \cite{nojiri2022barrow} demonstrated that the Barrow entropic dark energy DE model is identical to the generalized HDE, where the holographic cut-off is defined by the particle horizon and derivative and the future horizon and derivative. Such a generalized version of HDE, introduced by \cite{odintsov2019unification}, leads to interesting phenomenology, both from inflation and dark energy perspective \cite{nojiri2021different}. In this study, such a generalized version of HDE would be referred to as Nojiri-Odintsov holographic dark energy (NO-HDE). We will consider, inspired by \cite{khurshudyan2016holographic}, this highly generalized model of holographic dark energy with the Nojiri-Odintsov cut-off is defined as follows \cite{khurshudyan2016holographic}:
\begin{equation}
    \rho_{NO-HDE}=\frac{3c^2}{L^2}
\end{equation}
where,
\begin{equation}
    \frac{c}{L}=\frac{1}{L_f}\left[\alpha_0+\alpha_1 L_f+\alpha_2 L_f^2\right]
\end{equation}
where $L_f$ is the future horizon and is defined as
\begin{equation}
    L_f=a \int_{t}^{\infty}\frac{dt}{a}
\end{equation}
Following Eq. (\ref{Rh}), in the present case,
\begin{equation}\label{Lf}
L_f=\frac{t}{-1+\gamma}+C_1 t^\gamma
\end{equation}
Considering a holographic connection of $f(G)$ gravity with the highly generalized version of HDE, referred to as NO-HDE, we will consider $\rho_{NO HDE}=\rho_G$. Hence, we have the following differential equation:
\begin{equation}\label{ODENO}
\frac{1}{4} \left(-4 f(t)-t f'(t)+\frac{t
\left(5 f'(t)+t f''(t)\right)}{-1+\gamma }\right) = \frac{3 \left(\alpha_0+\alpha_1 \left(C_1 t^{\gamma }+\frac{t}{-1+\gamma }\right)+\alpha_2 \left(C_1
t^{\gamma }+\frac{t}{-1+\gamma }\right)^2\right)^2}{\left(C_1 t^{\gamma }+\frac{t}{-1+\gamma }\right)^2}  
\end{equation}
\begin{figure}[h]
\centering
\begin{minipage}{1.0\textwidth}
\centering
\includegraphics[width=0.55\linewidth, height=0.20\textheight]{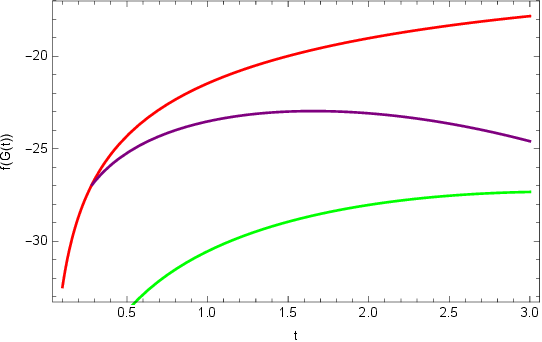}
\caption{The evolution NO HDE based reconstructed $f(G)$ gravity obtained by numerically solving differential equation (\ref{ODENO}). We have taken $\alpha_0 = 0.1$, $\alpha_1 = 0.2$ and $\alpha_2 = 0.3$. Red, green, and purple lines correspond to $\gamma=0.82,~0.85$ and $0.88$ respectively.}
\label{freconstNO}
\end{minipage}
\begin{minipage}{1.0\textwidth}
\centering
\includegraphics[width=0.55\linewidth, height=0.20\textheight]{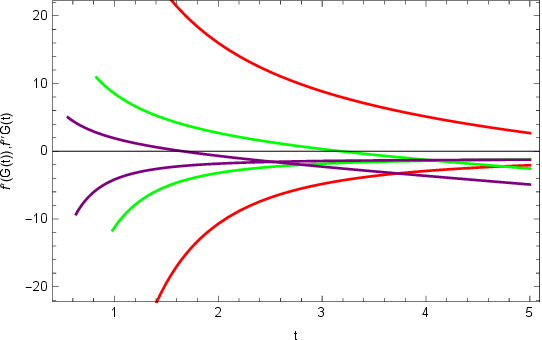}
\caption{The evolution of the derivates of NO HDE based reconstructed $f(G)$ gravity obtained by numerically solving differential equation (\ref{ODENO}). We have taken $\alpha_0 = 0.1$, $\alpha_1 = 0.2$ and $\alpha_2 = 0.3$. Red, green, and purple lines correspond to $\gamma=0.82,~0.85$ and $0.88$ respectively.}
\label{fdashreconNO}
\end{minipage}
\end{figure}
The above differential equation is solved numerically, and the solutions are plotted. In Fig. 7, we plotted the reconstructed $\rho_G$ as a cosmic time $t$ function with a holographic connection through the Nojiri-Odintsov cutoff. In Fig. \ref{freconstNO}, we observed that $f(G(t)) \rightarrow -\infty$ as $t \rightarrow 0$. From Eq. (\ref{G}), we note that $G \rightarrow 0$ and $t \rightarrow \infty$. Combining all these, we can say that the holographically connected $f(G)$ gravity satisfies a sufficient condition for a realistic model. In Fig. 8 we have plotted $f_G(t)$ and $f_{GG}(t)$. The upper panel contains  $f_G(t)$ and the lower panel  $f_{GG}(t)$. It is discerned that $f_G(t)$ is decreasing with $t$ and $f_{GG}(t)$ is becoming negative as a consequence. In Fig. 7, we have plotted the reconstructed density for $f(G)$ gravity, which shows the validity of WEC as well. The density is staying at positive level and this implies realistic reconstruction of $f(G)$ density through NO HDE. Furthermore, the reconstructed $\rho_G$ is increasing with the evolution with the universe. It is consistent with the DE-dominated late-time universe.
\begin{figure}[ht] 
  \begin{minipage}[b]{0.55\linewidth} 
    \includegraphics[width=.75\linewidth]{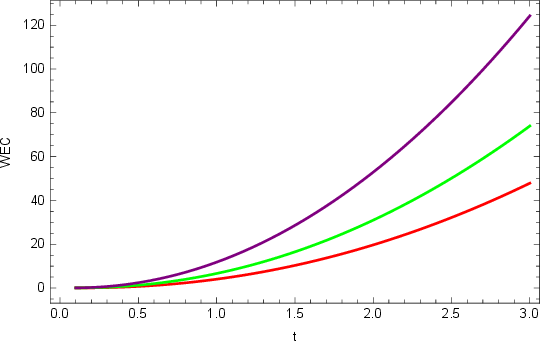} 
    \caption{$\rho_G \ge 0$.}\label{EC}  
  \end{minipage} 
  \begin{minipage}[b]{0.55\linewidth}
    \includegraphics[width=.75\linewidth]{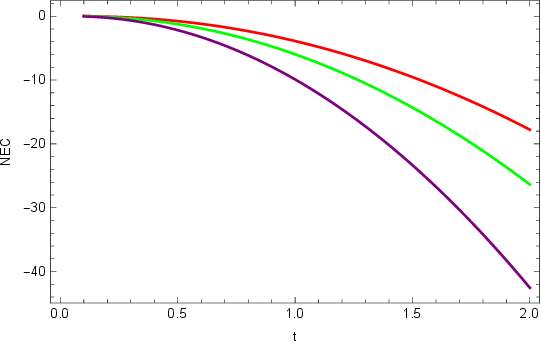} 
    \caption{$\rho_G+p_G \le 0$.}
  \end{minipage} 
  \begin{minipage}[b]{0.55\linewidth}
    \includegraphics[width=.75\linewidth]{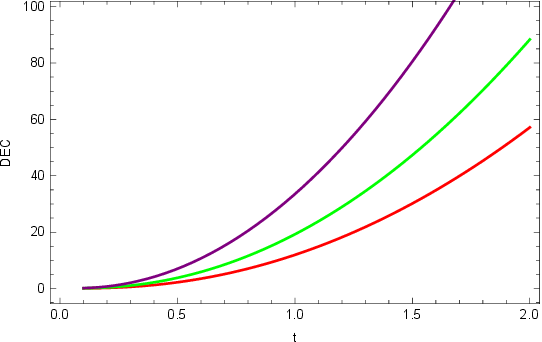} 
    \caption{$\rho_G-p_G \ge 0$}
  \end{minipage}
  \hfill
  \begin{minipage}[b]{0.55\linewidth}
    \includegraphics[width=.75\linewidth]{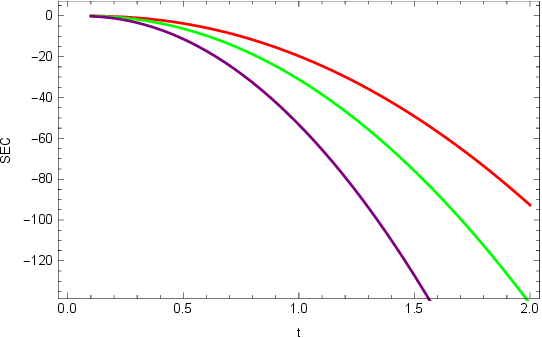} 
    \caption{$\rho_G+3p_G \le 0$}
  \end{minipage} 
\end{figure}
In Figs. 9, 10, 11 and 12, we have demonstrated the ECs for the reconstructed $\rho_G$ based on NO HDE. Fig. 10 shows that NO-HDE reconstructed $f(G)$ violates the NEC condition and this clearly indicates the possibility of EoS parameter to be less than $-1$. Near $t\approx 0$, the $\rho_G +p_G\rightarrow 0$ and hence it is close to the cosmological constant near the beginning of the universe. Fig. 11 shows the validity of DEC and Fig. 12 shows violation of the SEC. This ensures the accelerated expansion of the universe under this NO HDE reconstructed $f(G)$ gravity.

\section{Relation to observational bounds}
\cite{bamba2012dark} took a comprehensive approach to comparing alternate gravity and dark energy models to general relativity by cosmography.
They \cite{bamba2012dark} provided examples as well as evidence that good data analysis of huge data samples can eliminate parameter degeneration. \cite{bamba2012dark} came to the conclusion that the current work is a first step toward obtaining more accurate restrictions on the universe's expansion history, and that adding more observational datasets could aid in this process. Chevallier-Polarski-Linder parametrization was used in \cite{kumar2014observational} to demonstrate that, for dark energy EoS, current $/$ future observations favor the CDM scenario over the WDM. The EoS parameter's current value was restricted by $-1.06_{-0.13}^{+0.11}$ using observational data sets from $Planck + WMAP9 + WiggleZ$, BAO, and SNLS3.  In a more recent research by \cite{kumar2014observational}, the $95\%$ constraints imposed by Planck data in conjunction with other astrophysical measurements \cite{ade2016planck} constrain the present value of DE EoS by observations to be about $-1.019_{-0.080}^{+0.075}$. In the current section, we use the aforementioned observational bounds on the EoS parameter to compare the values of the reconstructed EoS parameter for various model parameter combinations. We consider Eq. (\ref{wG}) to study the validity of our reconstruction model against observational bounds presented in the above-mentioned studies and present the values of $w_G$ for $z=0$ presented in Table \ref{table}. 
\begin{table}[h!]
\caption{Current values $(z=0)$ of $w_{DE}$ based on Eq. (\ref{wG}). Values are compared with current EoS values $-1.06_{-0.13}^{+0.11}$ from observational data sets from SNLS3, BAO and
Planck + WMAP9 + WiggleZ available in \cite{kumar2014observational}.}
\begin{center}
\begin{tabular}{||c c c c||}
 \hline
 $\mathbb{C}$ & $\gamma$ & $\Delta$ & $w_G$ \\ [0.5ex] 
 \hline\hline
 0.8 & 2 & 0.1 & -0.9439 \\ 
 \hline
 0.9 & 1.8 & 0.5 & -0.9651 \\
 \hline
 0.7 & 1.5 & 0.9 & -1.0131 \\
 \hline
 0.8 & 2 & 1.5 & -1.0064 \\
 \hline
 0.8 & 1.8 & 1.2 & -1.0140 \\ [1ex] 
 \hline
\end{tabular}
\label{table}
\end{center}
\end{table}
The present values of the EoS parameter are consistent with observational data for various combinations of model parameter values, as seen from Table \ref{table}. It may be noted that we have taken $C_1=1.5, C_2=1.3, C_3=0.1, a_0=0.9$ and $\rho_{m0}=0.32$ for all the rows of Table \ref{table}.
\section{Conclusions}
The work presented in the preceding sections considers a holographic reconstruction approach for $f(G)$ gravity. There are two stages to the reconstruction process. In the study's first phase, we examined a relationship between $f(G)$ gravity and BHDE. 
The Barrow holographically reconstructed $f(G)$ gravity is plotted for a change in $\Delta$ in Fig. \ref{fig1}, where we graphically displayed the reconstructed $f(G)$. We previously explained that the BHDE density returns the standard holographic dark energy under the imposition $S \propto A \propto L^2$ \cite{saridakis2020barrow, wang2017holographic}. The condition $\rho_{DE}L^4 \le S$ establishes the standard holographic dark energy, where $L$ is the horizon length. We have taken this into account and restricted ourselves to $\Delta \in (0,1)$ when charting the reconstructed $f(G)$ in Fig. \ref{fig1}, where the evolutionary pattern of $f(G)$ has been evaluated versus redshift $z$. Depletion of energy density as the Universe expands is represented as a violation of the NEC. In this instance, the ECs of the current model are displayed in Figs. 3, 4, 5, and 6. Figure 4 illustrates the NEC violation. This implies that $w_G<-1$ since $\rho_G+p_G<0$. This suggests that the EoS parameter may exhibit phantom behavior. The SEC violation, $\rho_G+3 p_G <0$, which implies $w_G<-1/3$, is depicted in Fig. 6. This aligns with the universe's accelerating expansion. WEC's validity is demonstrated in Fig. 3, and DEC is likewise satisfied in Fig. 5. Using the most generalized version of holographic dark energy introduced by \cite{nojiri2007modified} and \cite{nojiri2017covariant}. To assess the validity of our reconstruction model against the observational bounds presented in the aforementioned studies, we used Eq. (\ref{wG}) in the final phase of the study. We also presented the values of $w_G$ for $z=0$ in Table \ref{table}. We found that the current values of the EoS parameter are consistent with observational data for a variety of model parameter value combinations, as shown in Table \ref{table}.

It has been shown in reference \cite{odintsov2019unification} that the coincidence problem in generalized holographic dark energy can be solved. According to \cite{odintsov2019unification,odintsov2023holographic}, particle and future event horizons, as well as their derivatives, have an impact on the holographic cut-off (LIR). We examined $\rho_{NO HDE}=\rho_G$ in light of a holographic relationship of $f(G)$ gravity with the highly generalized form of HDE, known as NO-HDE. As a result, we created a differential equation and used numerical methods to solve it. We used the Nojiri-Odintsov cutoff to represent the reconstructed $\rho_G$ as a cosmic time $t$ function with a holographic link in Figure 7. We saw that $f(G(t)) \rightarrow -\infty$ as $t \rightarrow 0$ in Fig. \ref{freconstNO}. We observe that $G \rightarrow 0$ and $t \rightarrow \infty$ from Eq. (\ref{G}). By combining all of them, we may conclude that the holographically connected $f(G)$ gravity satisfies a sufficient criterion for a realistic model. We have shown $f_G(t)$ and $f_{GG}(t)$ in Fig. 8. $f_G(t)$ and $f_{GG}(t)$ are found in the upper and lower panels, respectively. It can be shown that $f_G(t)$ decreases with $t$, and as a result, $f_{GG}(t)$ turns negative.

While concluding, let us make some final comments. In the current study we developed a connectivity between the most generalized holographic dark energy density \cite{nojiri2017covariant} and a particular case of it in the form of BHDE with $f(G)$ gravity. The reconstructed models were tested for the energy conditions. The outcomes are in consistency with \cite{koussour2022quintessence} and \cite{jawad2013reconstruction}. In future, we aspire to focus of the thermodynamics under such reconstruction scheme. 

\section*{Acknowledgements}
The insightful comments from the anonymous reviewer are thankfully acknowledged. The author sincerely acknowledges the hospitality of the Inter-University Centre for Astronomy and Astrophysics (IUCAA), Pune, India, during a scientific visit from December 2024 to January 2025 under  visiting associateship. 

\section*{Conflict of interest}
The author hereby declares that this work has no conflict of interest.

\end{document}